\documentclass[%
 reprint,
 amsmath,amssymb,
 aps,
]{revtex4-2}

\usepackage{graphicx}
\usepackage{dcolumn}
\usepackage{bm}
\usepackage[dvipsnames]{xcolor}
\usepackage{subcaption}
\begin{document}

\preprint{APS/123-QED}

\title{Trans-Planckian censorship and other swampland bothers
addressed in warm inflation}

\author{Arjun Berera}
    \email{ab@ph.ed.ac.uk}
\author{Jaime R. Calder\'on}%
 \email{jaime.calderon@ed.ac.uk}
\affiliation{%
 \\
 School of Physics and Astronomy, University of Edinburgh, Edinburgh, EH9 3FD, United Kingdom
}%





\begin{abstract}
The implications of the recently proposed Trans-Planckian censorship conjecture (TCC) are analyzed in the context of warm inflation. It is found that for a single-stage accelerated expansion the constraints imposed by the censorship are roughly the same as for cold inflation. Next, we study how a two-stage inflationary expansion with an intermediate radiation-dominated era can alleviate the bounds imposed by the censorship. For a demonstrative toy model we found $r<10^{-23}$, but can be $r<10^{-5}$ for a weaker form of TCC for the later stages of expansion, while still satisfying the other swampland conditions.
\end{abstract}

\maketitle



\section{\label{sec:intro}Introduction}

Inflation remains to this day as the most plausible mechanism to explain the large scale structure of the universe and the anisotropies in the Cosmic Microwave Background \cite{Brout:1977ix,Mukhanov:1981xt}. Furthermore, it single-handedly solves the drawbacks of the standard Big Bang cosmology, in particular, the horizon problem \cite{Guth:1980zm,Fang:1980wi,Linde:1982uu,linde1982new}. However, there are a few shortcomings with the inflationary paradigm. For instance, from a field theory perspective it has been proven difficult to keep radiative corrections to the effective potential under control so that the required flatness of the potential do not get spoiled (the \emph{eta problem}). Another complication is related to the fact that if inflation lasted long enough, present-day perturbative modes could be traced back to sub-Planckian wavelengths during inflation, raising questions about the validity of its predictions, since physics at those scales is not understood. This is known as the \emph{Trans-Planckian problem} \cite{Brandenberger:2000wr,Martin:2000xs}.

There are many approaches to tackle this problem, like considering modified dispersion relations at trans-Planckian (TP) frequencies, or introducing a new physics hypersurface such that perturbative modes do not evolve at said scales, although some residues of that era are predicted in the power spectrum \cite{Brandenberger:2009rs,Brandenberger:2012aj}. Further changes can be expected from the selection of initial states different to the Bunch-Davies vacuum, like the $\alpha-$vacua \cite{Shankaranarayanan:2002ax,Danielsson:2002qh,Goldstein:2002fc}, a standard practice to deal with the past-incomplete nature of inflation. In this sense, a different route is to consider non-singular bouncing cosmologies or other pre-inflationary models where one can keep observational relevant scales far from sub-Planckian wavelengths, yielding to negligible corrections from TP physics \cite{Brandenberger:2011gk}. These are considered past-complete extensions to the inflationary paradigm, such that the perturbations can be safely assumed to begin in the standard Bunch-Davies state \cite{Cai:2017pga}.

Against this background, it has been conjectured that any model susceptible to the TP problem belongs to the swampland \cite{Bedroya:2019tba,Bedroya:2019snp}. This has been coined the \emph{Trans-Planckian censorship conjecture} (TCC). It restricts long-lived de Sitter states but it does allow
for short-lived states that do not last long enough for
a TP mode to cross the horizon. In particular, the
TCC states that a meta-stable de Sitter state can exist
for a time $t \le (1/H) \ln(Mp/H)$. This restricts the energy scale of a phase of inflation to be fairly low, hence placing a bound on the tensor-to-scalar ratio of $r < {\cal O}(10^{-30})$, as shown in \cite{Bedroya:2019tba}.
Even though there are models that satisfy this constraint \cite{Tenkanen:2019wsd,Kamali:2019xnt,Schmitz:2019uti,Kadota:2019dol}, it has been also pointed out that the bound is less demanding for bouncing cosmologies and other pre-inflationary scenarios, basically for the same reasons that they are less sensitive to the TP problem \cite{Bedroya:2019tba,Cai:2019hge}. Along those lines, other scenarios like a multi-stage inflation \cite{Mizuno:2019bxy}, excited initial states \cite{Brahma:2019unn} or non-standard expansion histories \cite{Dhuria:2019oyf,Torabian:2019zms} can also mitigate the TCC constraints.

Notice however that the TCC comes in addition to the other swampland criteria, which bounds the field variation during inflation (the distance conjecture) and the shape of the scalar potential (the de Sitter conjecture). The swampland conditions emerge from the difficulty in constructing
a de Sitter space from string theory, at present the most developed
theory for quantum gravity. However, there are still
many fundamental issues to understand in string theory.
One viewpoint could be that inflation is very successful
in explaining observations and so the swampland conditions demonstrate the phenomenological inconsistency of string theory, at least
in its present form.
As such, one
could simply sidestep the problems emerging from string theory
and focus on the phenomenological success of inflation.
Alternatively, string theory has come the closest to
any idea so far in realizing a consistent theory of quantum
gravity. Thus, another viewpoint would be that
it is interesting to explore the extent of the consistency
that can be achieved between inflation and its realization
from string theory and its swampland conditions. This is
the viewpoint we will adopt for this paper.

Under such circumstances, warm inflation (WI) \cite{Berera:1995wh,Berera:1995ie} presents a unique set of features that allows it to overcome these constraints \cite{Motaharfar:2018zyb,Rasouli:2018kvy,Das:2018hqy,Das:2018rpg,Bastero-Gil:2018yen,Kamali:2019xnt}. The reason is twofold: firstly, the inflaton is considered to dissipate energy into a radiation heat bath. Thus, one can get the necessary amount of inflation with a small background field variation, provided that there is enough dissipation. Moreover, this also allows for steeper potentials in comparison to the standard picture (\emph{cold inflation}) \cite{BasteroGil:2009ec}, so that the dS conjecture can be readily satisfied while solving or at least mitigating the eta problem. In fact, the swampland conditions were stated in \cite{Berera:2004vm} more than a decade ago, of course not in those terms, but nevertheless as conditions necessary for a consistent high-energy model of warm inflation.
Secondly, not only dissipation is accounted for, but also the effects of a random noise term in accordance with a fluctuation-dissipation theorem. This determines the statistical properties of the inflaton fluctuations, which are now thermal. Consequently, warm inflation is conceptually more robust than the standard picture regarding the initial state of perturbations, since the observed fluctuations are already classical during inflation. Indeed, all fluctuations in warm inflation that become relevant to density perturbations are created while modes are in the horizon before crossing and at wavenumber scales well below TP scales. Thus, vacuum fluctuations are suppressed and play a small or no role in the primordial density fluctuations. However, it should be noticed that the TCC would still apply in this scenario, even though the role of vacuum fluctuations is not relevant.

In this paper we will explore the consequences of the TCC for the warm inflation scenario. We will show that the constraints imposed on the energy scale in warm inflation are roughly the same as for cold inflation. Subsequently, we will explore how these constraints can be mitigated in a multi-stage inflationary scenario with an intermediate radiation dominated phase, a setup that can be naturally produced by warm inflation from a field theory perspective. We analyze this for generic realizations of the model and then for a toy model. Finally, we point out that if one relaxes the TCC for later stages, the bound on the amplitude of tensor perturbations can be alleviated by several orders of magnitude.

\section{Warm Inflation}

\subsection{Background dynamics}

Warm inflation is an appealing alternative to the standard picture, both conceptually and phenomenologically. It can be seen as a generalization of the standard picture where the interaction of the field with radiation degrees of freedom cannot be neglected during inflation. An extra bonus of this assumption is the absence of a reheating phase, since there is a smooth transition between inflation and the radiation-dominated (RD) era.

Dissipation of energy is readily accounted for through a friction-like term in the equation of motion of the inflaton, such that
\begin{equation}
    \ddot{\phi} + (3H + \Upsilon)\dot{\phi} = -V_{,\phi},
\end{equation}
where $\Upsilon$ is known as the dissipative coefficient. It is convenient to introduce a dissipative ratio $Q = \Upsilon/3H$, which quantifies the importance of dissipation in comparison to the expansion rate. On the other hand, the evolution of radiation is determined through the conservation of the energy-momentum tensor, which renders the continuity equation
\begin{equation}
    \dot{\rho}_r + 4H \rho_r = \Upsilon \dot{\phi}^2.
\end{equation}
Because of this extra friction term, the slow-roll conditions in warm inflation are more relaxed than their cold inflation counterparts. As in that scenario, one expects the field acceleration $\ddot{\phi}$ to be small in comparison to the other terms, so the kinetic energy does not surpass the potential energy, effectively ending inflation too soon. This can be checked through the slow-roll parameters, which in warm inflation are generalized as follows
\begin{equation} \label{slow-roll}
    \epsilon = \frac{M_p^2}{2(1+Q)}\left(\frac{V_{,\phi}}{V}\right)^2, \hspace{1cm} |\eta| = \frac{M_p^2}{1+Q}\left|\frac{V_{,\phi \phi}}{V}\right|,
\end{equation}
where $M_P$ is the reduced Planck mass and $\epsilon \simeq \epsilon_H \equiv -\dot{H}/H^2$ during slow-roll. Naturally, the end-of-inflation condition remains the same as in the standard picture, i.e., $\epsilon_H= 1$. From this, it becomes clear that the distance and dS swampland constraints are more easily overcome in warm inflation, as the field can move less due to dissipation, which also allows for steeper potentials.

Finally, it is worth mentioning that warm inflation asks for further consistency checks. Arguably, the most relevant is that $T>H$, which can be interpreted as the requirement that the microscopic dynamics should exceed the expansion rate in order to maintain a thermal state. Likewise, in quantum field theory realizations of warm inflation more stringent
consistency conditions also include that the time scales of all relevant
microphysical processes are faster than the Hubble rate, $\Gamma>H$,
where $\Gamma$ represents any microphysical decay and/or scattering
rates in the system.

\subsection{Computation of Perturbations}

Warm inflation presents fundamental differences in the study of perturbations in comparison to cold inflation. For starters, the presence of other fields, namely radiation, also generate density fluctuations that contribute to the curvature power spectrum. In this way, the comoving curvature perturbation can be written as
\begin{equation}
    {\cal R} = -\frac{H}{\rho + p} \Psi_T = -\frac{H}{\rho + p} (\Psi_{\phi} + \Psi_r),
\end{equation}
where $\rho$ and $p$ are the total energy density and pressure respectively, and $\Psi_T$ denotes the total momentum perturbation in the spatially-flat gauge. It has been shown numerically \cite{BasteroGil:2011xd} and analytically \cite{Bastero-Gil:2019rsp} that at horizon crossing radiation and momentum perturbations are related by
\begin{equation}
    \Psi_r \simeq Q \Psi_{\phi},
\end{equation}
rendering a curvature perturbation of the form
\begin{equation}
    {\cal R} = -\frac{1}{2M_p^2 \epsilon}(1+Q)\Psi_{\phi} \simeq \frac{H}{\dot{\phi}}\delta\phi,
\end{equation}
where we have used $\Psi_{\phi} = -\dot{\phi} \delta \phi$ and the slow-roll approximation, which implies $\rho + p \simeq (1+Q)\dot{\phi}^2$. Consequently, and similarly to cold inflation, the curvature power spectrum reads
\begin{equation}
    \Delta_{\cal R}^2 = \left(\frac{H}{\dot{\phi}}\right)^2 \Delta_{\delta\phi}^2.
\end{equation}
Nevertheless, the amplitude of the spectrum, $\Delta_{\delta\phi}^2$, is quite different to its standard picture counterpart, as already mentioned. Indeed, dissipation induces the field perturbation $\delta\phi$ to satisfy a Langevin-like equation, with a fluctuation-dissipation relation determining the statistical properties of the fluctuations. In this way, we get \cite{Ramos:2013nsa,Bartrum:2013fia}
\begin{equation}
    \Delta_{\delta\phi}^2 = \left(\frac{H_*}{2\pi}\right)^2 \left[1+2n_* + \frac{T_*}{H_*}\frac{2\pi\sqrt{3}Q_*}{\sqrt{3+4\pi Q_*}}\right],
\end{equation}
where $n_*$ denotes the statistical distribution of the perturbative modes, and each background quantity is evaluated at horizon crossing.

Finally, the observables linked to perturbations, the spectral index and tensor-to-scalar ratio, are defined by
\begin{equation}
    n_s -1 = \frac{d \ln \Delta_{\cal R}^2}{dN}, \hspace{1cm} r = \frac{\Delta_T^2 (k_0)}{\Delta_{\cal R}^2 (k_0)},
\end{equation}
where $\Delta_T^2$ is the power spectrum of tensor perturbations given by
\begin{equation}
    \Delta_T^2 = \frac{8}{M_p^2} \left(\frac{H}{2\pi}\right)^2,
\end{equation}
and $\Delta^2_{\cal R}(k_0) = 2.2 \times 10^{-9}$, where $k_0 = 0.002\ \textrm{MPc}^{-1}$ is a pivot scale \cite{Akrami:2018odb}. On the other hand, in the case of strong dissipation the spectral index can be approximated by \cite{Bastero-Gil:2019rsp}
\begin{equation}
    n_s \approx 1+ \frac{3}{4}(2\eta - 6 \epsilon) + \left(\frac{5}{4} - \frac{1}{\sqrt{3\pi Q}}\right)\theta,
\end{equation}
where $\theta= d \ln(1+Q)/dN$ is a slow-roll parameter describing the evolution of dissipation.

\subsection{Swampland constraints}

\subsubsection{The (refined) distance conjecture}

This conjecture was motivated by the difficulty to embed large-field inflationary models into string theory \cite{Palti:2019pca,klaewer2017super}. Then, it is inferred that for large distances $d$ in field space of the effective theory, there is an infinite tower of states with mass
\begin{equation}
    m \sim M_p e^{-\alpha d},
\end{equation}
with $\alpha \sim {\cal O}(1)$. Hence, the distance conjecture effectively sets an upper bound for the energy scale of inflation, such that
\begin{equation}
    \Lambda_{dc} \equiv A e^{-\alpha \Delta\phi/M_p} M_p > E_{inf},
\end{equation}
where $E_{inf}$ is defined in terms of the tensor-to-scalar ratio as
\begin{equation}
    E_{inf} \simeq V^{1/4} \simeq 7.6 \times 10^{-3} \left(\frac{r}{0.1}\right)^{1/4} M_p.
\end{equation}
There is not a fixed value for the parameters $A$ and $\alpha$, however, one can approximate this condition as \cite{Das:2018rpg,Das:2019hto}
\begin{equation}\label{dcond}
    \frac{\Delta\phi}{M_p} < \Delta \sim {\cal O}(1).
\end{equation}
This limit comes into tension with the Lyth bound, which for warm inflation is given by
\begin{equation}
    \frac{\Delta\phi}{M_p} = \int dN \sqrt{\frac{r}{8}} \left[1+2n_*+\frac{T_*}{H_*}\frac{2\pi\sqrt{3}Q_*}{\sqrt{3+4\pi Q_*}}\right]^{-1/2}
\end{equation}
Notice that this is a more severe problem in the case of cold inflation, since the Lyth bound favours large field models, whereas the distance conjecture does the opposite. The possibility of having strong dissipation and hence a small field motion brings warm inflation in consistency with these bounds.

\subsubsection{The de Sitter conjecture}

The difficulty of constructing a de Sitter vacua from string theory has hinted that any EFT that does present that feature actually belongs to the swampland \cite{obied2018sitter,garg2018bounds,ooguri2007geometry}. In this way, it has been posit that the scalar potential of an EFT coupled to gravity must satisfy either
\begin{equation}
    |\nabla V| \ge \frac{c}{M_p} V, \hspace{0.6cm} \textrm{or} \hspace{0.6cm} \min(\nabla_i\nabla_j V) \le -\frac{c'}{M_p^2} V,
\end{equation}
where $c$ and $c'$ are positive constantes of order $1$. In this sense, it is clear from Eq. \eqref{slow-roll} that warm inflation can generically satisfy these conditions, even more so if $Q\gg 1$.

\subsection{Trans-Planckian Censorship Conjecture in Warm Inflation}\label{TCC1}

Recently, it has been proposed that any EFT consistent with string theory should not lead to an expansion period with perturbation lengths that can be traced back to sub-Planckian scales \cite{Bedroya:2019snp,Bedroya:2019tba}. Then, in the string theory context, there should not exist a TP problem. In consequence, no TP mode can cross the horizon, which translates into
\begin{equation}
    \frac{l_P}{a_i} < \frac{1}{a_f H_f},
\end{equation}
where $l_P$ denotes the Planck length, $a_i\ (a_f)$ the scale factor at the start (end) of inflation, and $H_f$ is the Hubble parameter at the end of inflation. On the l.h.s we find the comoving length of the largest TP mode at the beginning of inflation and on the r.h.s the comoving horizon at the end of inflation. Rearranging terms, this is equivalent to
\begin{equation}\label{tcc1}
    e^{N_e} = \frac{a_f}{a_i} < \frac{M_p}{H_f}.
\end{equation}
Naturally, inflation needs to last long enough to solve the horizon problem, but as it can be seen, the TCC bounds the number of e-folds from above, imposing further constraints on the energy scale of inflation and the amplitude of tensor perturbations. Henceforth, we will specialize the analysis for the warm inflation case, which presents some minor differences in comparison to the material presented in \cite{Bedroya:2019tba}.

Firstly, in order to solve the horizon problem, the present comoving horizon has to be contained within the comoving horizon at the beginning of inflation, i.e.,
\begin{equation}
    \frac{1}{a_0 H_0} < \frac{1}{a_i H_i}.
\end{equation}
Once again, rearranging terms and conveniently introducing the scale factor at the end of inflation, the inequality above becomes
\begin{equation}
    \frac{1}{H_0} < \frac{a_0}{a_f}\frac{a_f}{a_i}\frac{1}{H_i} \Longleftrightarrow \frac{1}{H_0} < \frac{T_f g_*^{1/3} (T_f)}{T_0 g_*^{1/3} (T_0)} e^{N_e} \frac{1}{H_i},
\end{equation}
or equivalently,
\begin{equation}\label{tcc11}
    \frac{1}{H_0} < \frac{T_f}{T_0} e^{N_e} \frac{1}{H_i}\Longleftrightarrow \frac{H_i}{T_f}\frac{T_0}{H_0} < e^{N_e},
\end{equation}
where we have assumed that the ratio between the (cubic root) number of degrees of freedom at the end of inflation and at present day is of order one. Thus, Eqs. \eqref{tcc1} and \eqref{tcc11} imply
\begin{equation}\label{tcss}
    \frac{T_0}{H_0} < \frac{T_f}{H_f} \frac{M_p}{H_i}.
\end{equation}
Finally, assuming slow-roll and a rapid thermalization such that $\rho_r \propto T^4$,we have got that \footnote{We have omitted the proportionality factor in $\rho_r(T)$. For the sake of concreteness, we have chosen $g_* = 228.75$, corresponding to the MSSM. Other choices change the numerical values by a factor of ${\cal O}(1)$ at most.}
\begin{equation}\label{th}
    \frac{T_f}{H_f} \simeq \left[\frac{9}{2} \frac{Q_f}{1+Q_f}\right]^{1/4} \frac{M_p}{V_f^{1/4}}.
\end{equation}
In this context, the strong dissipative regime is the most interesting, since it helps to satisfy more easily the swampland criteria. Thus, the TCC constraint for warm inflation is
\begin{equation}
    V_i^{1/2} V_{f}^{1/4} < 5 \times 10^{-30} M_p^3,
\end{equation}
where we have used that $T_0/H_0 \approx 1.7 \times 10^{29}$. Furthermore, since $V_i > V_f$, we can find a bound for the energy scale at the end of inflation of
\begin{equation}
    V_f^{1/4} < 1.7 \times 10^{-10} M_p \sim 4 \times 10^8\ \textrm{GeV},
\end{equation}
in agreement with \cite{Das:2019hto}. The main difference with the expression found in \cite{Bedroya:2019tba} is due to the fact that in warm inflation there is no need to consider a reheating phase, and that, depending on the amount of dissipation, the Hubble parameter at the beginning of inflation could be up to two orders of magnitude larger than its value at the end of inflation. Consequently, there is a stringent constraint on the tensor-to-scalar ratio as well.

\section{A Multi-Stage WI response to the TCC}

\begin{figure*}
\includegraphics[width=0.9\textwidth]{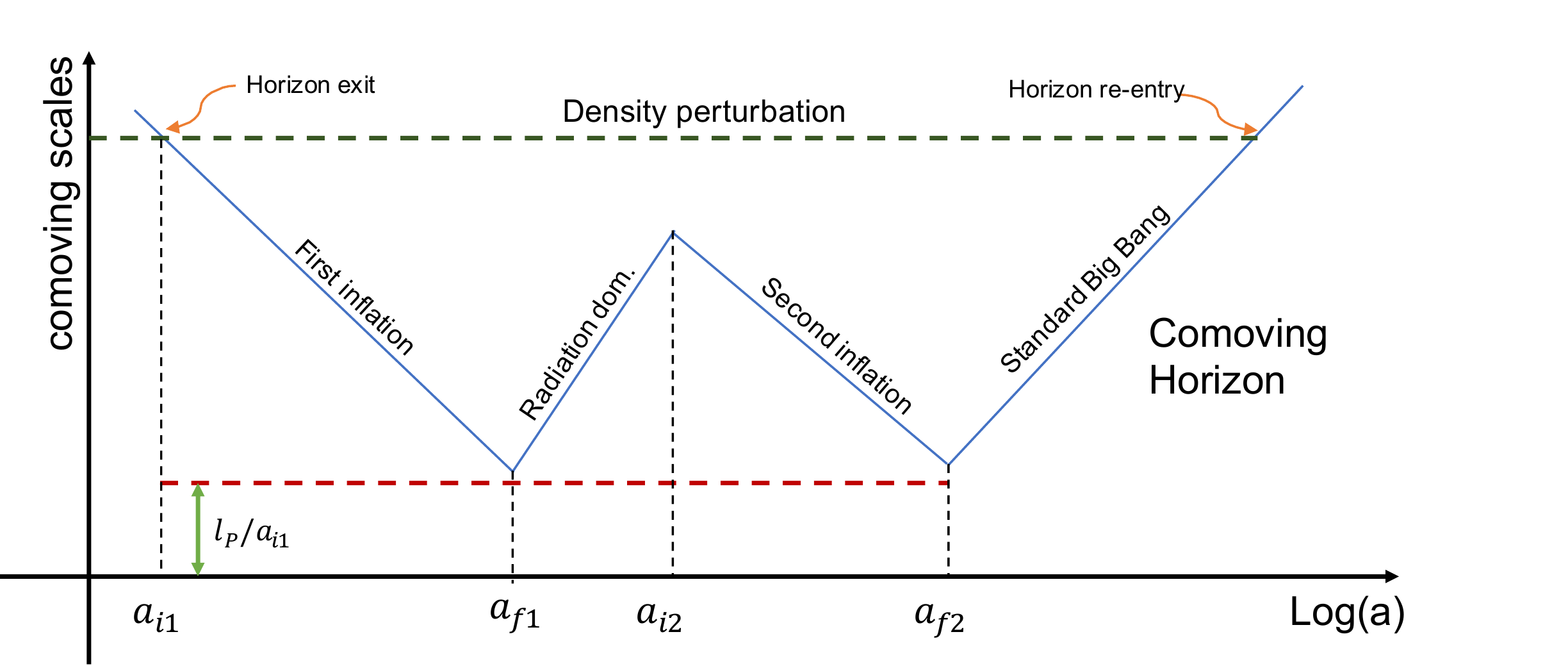}
\caption{\label{fig:wide} Pictorial illustration of the evolution of the comoving horizon for a two-stage warm-inflation scenario. Some relevant (comoving) scales are presented, like the largest TP mode at the beginning of inflation, or the perturbation crossing horizon at the same time. }
\end{figure*}

In this section we will explore how the TCC constrains the scale of inflation if the process takes place in multiple stages. For the sake of simplicity, a two-stage scenario will be considered, with an intermediate RD era, as illustrated on Fig. \ref{fig:wide}. In this sense, the first phase, during which the largest observable perturbation exits the horizon, happens at a higher energy scale. In contrast, the subsequent phases could happen at very low energies. Notice that this does not affect the amplitude of the (potentially) measurable tensor perturbations and thus, the Lyth bound does not apply for those low-energy periods. In this way, the main asset of the other phases is to mitigate the amount of inflation required for the first period, so that the more severe constraints are those enforced by the TCC. Along these lines, working in a warm inflation setup presents the further advantage/possibility of having intermediate radiation-dominated (RD) eras with smooth transitions in between, which would further alleviate the demands on the first period.

Scenarios such as the one described above are not strange to quantum field theory. Indeed, cosmological phase transitions can be a source for
small periods of inflation.
In \cite{Bartrum:2014fla} it was shown that dissipative effects
do occur generically in particle physics models.
The paper computed the dissipation coefficient for scalar
fields within the Standard Model and some of its common supersymmetric extensions like the minimal (MSSM) and next-to-minimal (NMSSM) supersymmetric SM.
These dissipative effects had significant impact on the evolution
of cosmological scalar fields, leading to friction and entropy
production. During phase transitions within these models it
was shown that periods of warm inflation would develop for a few
e-folds ${\cal O}(1-10)$,
even for the Electroweak transition in the Standard Model.
Moreover it has also been shown that scalar fields can
get trapped in a false vacuum by finite temperature effects,
leading to a short period ${\cal O}(1-10)$ of thermal
inflation \cite{Lyth:1995ka}.
As cosmological phase transitions are common in particle physics
models from the Standard Model to its various extensions,
multi-stages of inflation would not be unusual to expect.

There are well-known problems with plateau models
in cold inflation, since a slow-roll solution is not a phase space
attractor for most potentials of this type. This implies
inflation can only be triggered if the field is initially located
in the plateau and with sufficiently small velocity.
Moreover, this needs to hold on super-Hubble scales,
since nonlinear effects from both the scalar field and
the metric could prevent the start of inflation.
Thus irrespective if the conditions leading to inflation are satisfied
inside the Hubble-sized patch, if in the outer regions the
field value does not meet these requirements, the field and
space-time dynamics may eventually cause the field to exit
the slow-roll regime everywhere. As such, plateau
models require fine tuning of the initial conditions
for the field value, its velocity, and its degree of
homogeneity to realize inflation. It has even been
argued in \cite{Ijjas:2013vea,Ijjas:2014nta}
(although an alternative point of view was given
in \cite{Guth:2013sya,Linde:2014nna})
that the inflationary paradigm is at considerable
risk of falling if it requires plateau potentials.

In warm inflation, these fine tuning problems for plateau
potentials are not present. If the scalar field is
being governed by strong dissipation, $\Upsilon > H$, it was shown
in \cite{Berera:2000xz} that it damps fluctuations with physical
wavelength
$k < H$. This means smoothness of the initial pre-inflationary patch
need not require the Hubble scale $1/H$ but rather just $1/\Upsilon$.
At length scales bigger than that, the dissipation will damp
the modes and prevent nonlinear dynamics from becoming significant.
In \cite{Bastero-Gil:2016mrl} it was shown within
a quantum field theory SUSY model
that fluctuation-dissipation dynamics can be present during
the radiation-dominated pre-inflationary epoch that can thermalize
the state, with the inflaton field naturally becoming
localized with a flat plateau about the origin, thus setting
the necessary conditions for the onset of inflation.
Thus plateau models do not present an initial condition
fine tuning problem in warm inflation.

\subsection{Generic constraints}

Following the discussion above, we consider a potential which dominates the dynamics of the first expansion, which is effectively decoupled from the potential in charge of the second inflationary phase. Then, there are two TCC conditions that should be satisfied, namely,
\begin{eqnarray}
    e^{N_1} = \frac{a_{f1}}{a_{i1}}  & < & \frac{M_p}{H_{f1}},\label{tca1}\\
    e^{N_T} = \frac{a_{f2}}{a_{i1}}  & < & \frac{M_p}{H_{f2}},\label{tca2}
\end{eqnarray}
where $N_1$ is the number of e-folds of expansion during the first period, while $N_T$ denotes the total amount of expansion, including the RD phase. The first condition is the same as for a single-shot inflation, whereas the second comes from requiring that the largest TP mode ($\lambda = l_p$ at the beginning of inflation) does not cross the horizon at the end of the entire period.

Next, the horizon problem is dealt with in a similar fashion as in Section \ref{TCC1}, resulting in the condition
\begin{equation}\label{hor2}
    \frac{H_{i1}}{T_{f2}}\frac{T_0}{H_0} < e^{N_T}.
\end{equation}
Thus, combining Eqs. \eqref{tca2} and \eqref{hor2}, we get
\begin{equation}\label{tfa1}
    \frac{H_{i1}}{T_{f2}}\frac{T_0}{H_0} < e^{N_T} < \frac{M_p}{H_{f2}} \Longrightarrow \frac{T_0}{H_0} < \frac{T_{f2}}{H_{f2}}\frac{M_p}{H_{i1}}.
\end{equation}
This expression is analogous to Eq. \eqref{tcss} found for single-shot warm inflation. Consequently, the potentials at the different stages should satisfy
\begin{equation}
    V_{i1}^{1/2} V_{f2}^{1/4} < 5 \times 10^{-30}\ M_p^3,
\end{equation}
which in turn implies that $V_{f2}^{1/4} < 10^{-10}\ M_p$. Even though the upper bound on the low-energy inflation is the same for the single-phase scenario (both in cold and warm inflation), $V_{i1}$ can take much higher values, as long as Eq. \eqref{tca1} is satisfied. The conditions necessary for that are model-dependent, so we will leave that for the next subsection. However, there is still one generic point left to discuss. Once the perturbative mode of interest has crossed the horizon, it should not re-enter it before the second inflation starts. There is a possibility of that happening during the intermediate RD era if it lasts long enough. To avoid that, the comoving horizon at $t_{i2}$ should be smaller than the comoving horizon at the beginning of inflation, at $t_{i1}$, i.e.,
\begin{equation} \label{nret}
    \frac{1}{a_{i2}H_{i2}} < \frac{1}{a_{i1}H_{i1}} \Longleftrightarrow \frac{H_{i1}}{H_{i2}} < \frac{a_{i2}}{a_{i1}} = e^{N_1 + N_R},
\end{equation}
where $N_R$ denotes the amount of expansion during the intermediate RD phase. Consequently,
\begin{equation}\label{nr1}
    V_{i2}^{1/2} > V_{i1}^{1/2} e^{-(N_1+N_R)},
\end{equation}
so the \emph{no re-entry condition} sets a minimum scale for the second inflationary period, or an upper limit for the amount of expansion during the intermediate RD era. To quantify this, notice that the radiation energy density during RD is given by
\begin{equation}
    \rho_r (a) = \rho_r(a_{f1}) \left[\frac{a_{f1}}{a}\right]^4.
\end{equation}
On the other hand, the Stefan-Boltzmann equation together with Eq. \eqref{th} render
\begin{equation}
    \rho_r(a_{f1}) \simeq \frac{V_{f1}}{2},
\end{equation}
whereas the second inflationary phase will start roughly when $\rho_r \sim V_{i2}$, so that
\begin{equation}\label{nr2}
    \rho_r (a_{i2}) \simeq V_{i2} \simeq \frac{V_{f1}}{2}  \left[\frac{a_{f1}}{a_{i2}}\right]^4.
\end{equation}
This yields an expansion of
\begin{equation}\label{Nr}
    N_R = \ln \frac{a_{i2}}{a_{f1}} \simeq \frac{1}{4} \ln \frac{V_{f1}}{2V_{i2}}.
\end{equation}
Finally, combining \eqref{nr1} and \eqref{nr2}, one also finds that
\begin{equation}\label{Nrm}
    N_R < \frac{1}{2}\ln\frac{V_{i1}}{2V_{f1}}+N_1.
\end{equation}

\subsection{A toy model} \label{toy}

In order to explore a concrete example, we study a toy model with a potential
\begin{equation}
    V(\phi,\chi) = \lambda \phi^4 + \kappa M^4 \left(1-4 e^{-2\chi/M}\right),
\end{equation}
where $\phi$ drives the first inflationary expansion and $\chi$ the second one. Notice that the second term is a plateau potential and the large field approximation of the function $V(\chi) = \kappa M^4 \tanh^2\left(\chi/M\right)$.
This kind of potentials, which can be embedded into supergravity, have been studied in the context of $\alpha$-attractor inflation \cite{Kallosh:2013xya,Kallosh:2013hoa}, including double inflation realizations \cite{Maeda:2018sje}.

For each inflationary phase, dissipation will be modelled such that analytical calculations can be simplified as much as possible at this stage. However, there are many quantum field theory derived models of warm inflation \cite{Berera:1998px,Berera:1999ws,Bartrum:2013fia,Rosa:2018iff,Bastero-Gil:2016qru,Bastero-Gil:2019gao,Berghaus:2019whh}. A fundamental difference between those models and the one we will use in this work is the temperature dependence of the dissipative coefficient, which translates into a much richer phenomenology.

\subsubsection{First inflationary phase}

We consider the first phase to be dominated by a scalar field subject to a quartic potential. Dissipative effects are governed by a coefficient of the form $\Upsilon \propto \phi^2$. In this way, the dissipative ratio during slow-roll is given by
\begin{equation}
    Q_1 = \frac{\Upsilon_1}{3H} \simeq \frac{\Upsilon_0/M_p}{\sqrt{3\lambda}} = {\tt const.}
\end{equation}

On the other hand, the end of inflation condition happens when
\begin{equation}
    \epsilon = \frac{M_p^2}{2(1+Q_1)}\left(\frac{V_{\phi}}{V}\right)^2 = 1,
\end{equation}
which occurs when the field has a value
\begin{equation}\label{phf}
    \phi_f = \sqrt{\frac{8}{1+Q_1}}M_p.
\end{equation}
Likewise, the value of the field $N_1$ e-folds prior to the end of the first inflation can be readily found through
\begin{equation}
    N_1 \simeq -\frac{1}{M_p^2} \int_{\phi_i}^{\phi_f} d\phi\  \frac{V}{V_{,\phi}}(1+Q_1),
\end{equation}
yielding
\begin{equation}\label{phi}
    \phi_i = \sqrt{\frac{8(1+N_1)}{1+Q_1}}M_p.
\end{equation}
Notice that this is the value of the field at horizon crossing. Thus, the spectrum is fitted evaluating background quantities at this instant. Next, taking the ratio between the potential at the beginning and the end of this period, we find
\begin{equation}\label{N1}
    N_1 = \left(\frac{V_{i1}}{V_{f1}}\right)^{1/2} - 1,
\end{equation}
as it can be easily seen from Eqs. \eqref{phf} and \eqref{phi}. From the same equations, it is clear that the distance swampland condition in Eq. \eqref{dcond} is readily satisfied for $Q_1 \gg 1$.

Finally, invoking the TCC through Eq. \eqref{tca1} into the expression above, we conclude that
\begin{equation}\label{vif1}
    V_{i1}^{1/2} < V_{f1}^{1/2} \left[1+\ln \frac{\sqrt{3} M_p}{V_{f1}^{1/2}}\right].
\end{equation}
Then, for example, for $V_{i1}^{1/4} \sim 10^{-3}\ M_p$, one would need $V_{f1}^{1/4} > 2.34 \times 10^{-4} \ M_p$, just to avoid the crossing of TP modes. Likewise, $V_{i1}^{1/4} \sim 10^{-7}\ M_p$ induces a bound $V_{f1}^{1/4} > 1.64 \times 10^{-8}\ M_p$.

\subsubsection{Second inflationary phase}

For this phase, a second scalar field slowly rolling through a plateau potential drives the expansion of the universe. Physical processes like the dissipation of energy into radiation should happen at a rate comparable to a characteristic mass scale during that era. Locally, we have got that
\begin{equation}
    |V_{,\chi\chi}| = 16 \kappa M^4 e^{-2\chi/M} \sim m^2,
\end{equation}
and thus we will take
\begin{equation}
    \Upsilon_2 = \kappa^{1/2} M e^{-\chi/M}.
\end{equation}
This renders a dissipative ratio
\begin{equation}
    Q_2 = \frac{1}{\sqrt{3}}\frac{M_p}{M} \frac{e^{-\chi/M}}{(1-4 e^{-2\chi/M})^{1/2}} \simeq \frac{1}{\sqrt{3}}\frac{M_p}{M}e^{-\chi/M}.
\end{equation}
Hence, the end of inflation condition $\epsilon (\chi_f) = 1$ becomes
\begin{equation}
    32 \sqrt{3} \frac{M_p}{M} \left(\frac{e^{-2\chi_f/M}}{1-4 e^{-2\chi_f/M}}\right)^{3/2} \simeq 1,
\end{equation}
where once again, we have assumed strong dissipation. Then, the value of the field at the end of second inflation can be approximated by
\begin{equation}
    \chi_f \simeq \frac{M}{3} \ln \left(32 \sqrt{3}\frac{M_p}{M}\right).
\end{equation}
Likewise, the dissipative ratio at the end of inflation is
\begin{equation}
    Q_{f2} \simeq \left[\frac{1}{12\sqrt{2}} \frac{M_p}{M}\right]^{2/3}.
\end{equation}
Finally, the amount of inflation is
\begin{equation}
    N_2 \simeq -\frac{1}{M_p^2} \int_{\chi_i}^{\chi_f} d\chi \frac{V}{V_{,\chi}}(1+Q_2).
\end{equation}
It is convenient to make a change of variables such that $x = e^{-2\chi/M}$. The large field assumption ($\chi \gg M$) is equivalent to $x_i,x_f \ll 1$. Hence, the integral can be well approximated by
\begin{equation}\label{eqn2}
    N_2 \simeq \frac{M^2}{16 M_p^2} \left.\left[\frac{1}{2x^2} + \frac{M_p}{\sqrt{3}M} \frac{1}{x}\right]\right|_{x_f}^{x_i},
\end{equation}
where
\begin{equation}\label{xf2}
    x_f = \left[32\sqrt{3} \frac{M_p}{M}\right]^{-2/3}.
\end{equation}
Then, one can always choose values of $M_p/M$ consistent with strong dissipation, in particular $Q_{f2} \sim 100$, which yields $x_f = 1/9600$. The initial condition of the field  is determined by solving \eqref{eqn2}, which in turn fixes the starting energy scale.

Collecting and summarizing the results, it was found that the total amount of expansion $N_T = N_1 + N_R + N_2$ should be such that
\begin{equation}\label{tcst}
    \frac{H_{i1}}{T_{f2}} \frac{T_0}{H_0} < e^{N_T} < \frac{M_p}{H_{f2}},
\end{equation}
where the expansion during the first inflation, RD phase and second inflation are given by Eqs.  \eqref{N1}, \eqref{Nr} and \eqref{eqn2}, respectively. In addition, the TCC for the first phase should be satisfied separately by means of Eq. \eqref{vif1} and, finally, the amount of expansion during radiation domination is constrained by Eq. \eqref{Nrm}, which guarantees that the mode that crossed the horizon at $t_{i1}$ does not re-enter during the intermediate stage.

One can also consider a weaker form of TCC (in contrast to the \emph{strong} one presented above) where the TP modes from
the first inflation are not constrained by the conditions during the second one. If string theory is the correct high energy complete theory, then
the swampland conditions put restrictions in any low energy
effective theory on the types of
scalar potentials that are consistent with it. The restrictions
in particular focus on the de Sitter states that can emerge
from the potential and put restrictions on the slope, curvature
and lifetime of such states. The TCC in particular
implies such de Sitter states can only be metastable and
last for short durations. In general therefore, behind the
scalar potentials that one writes down for inflation
there would be some complicated string theory dynamics
from which they emerge. If for example a scalar potential
has two different regions in its field space which
realize de Sitter spaces, how these spaces are related
in the underlying string theory would not be immediately clear
from just the potential of the low-energy theory.
If they were somehow completely disjoint de Sitter spaces, then
the TCC would apply separately for each space. That
could mean TCC would place no retriction on any preexisting modes when
a particular de Sitter space emerges.

In this weaker version, the TCC conditions become
\begin{equation}\label{tcca2}
    e^{N_1}  <  \frac{M_p}{H_{f1}}, \hspace{1cm} e^{N_2}  <  \frac{M_p}{H_{f2}}.
\end{equation}
Naturally, both the horizon Eq. \eqref{hor2} and the no re-entry Eqs. \eqref{nret},\eqref{Nrm} conditions still hold. Hence,
\begin{eqnarray}
    \frac{1}{a_0 H_0} < \frac{1}{a_{i1}H_{i1}}  \Longleftrightarrow  \frac{1}{H_0} & < & \frac{1}{H_{i1}}\frac{T_{f2}}{T_0}e^{N_T}\nonumber\\& < & \frac{e^{N_R}}{H_{i1}}\frac{T_{f2}}{T_0} \frac{M_p^2}{H_{f1}H_{f2}},
\end{eqnarray}
where, in the last part, we have applied Eq. \eqref{tcca2} for the $N_1$ and $N_2$ bits of the exponential. As usual, rearranging terms, we get
\begin{equation}
    \frac{T_0}{H_0} \frac{H_{i1}H_{f1}}{M_p^2} e^{-N_R} < \frac{T_{f2}}{H_{f2}}.
\end{equation}
This is to be compared with Eq. \eqref{tfa1}, which shows that this weaker TCC imposes a much looser constraint on the potentials, manifested by the exponential term denoting the expansion during radiation domination.

\subsubsection{Numerical Results}

\begin{figure*}[htp!]
  \centering
  \begin{subfigure}[b]{0.49\textwidth}

    \includegraphics[width=1.05\textwidth]{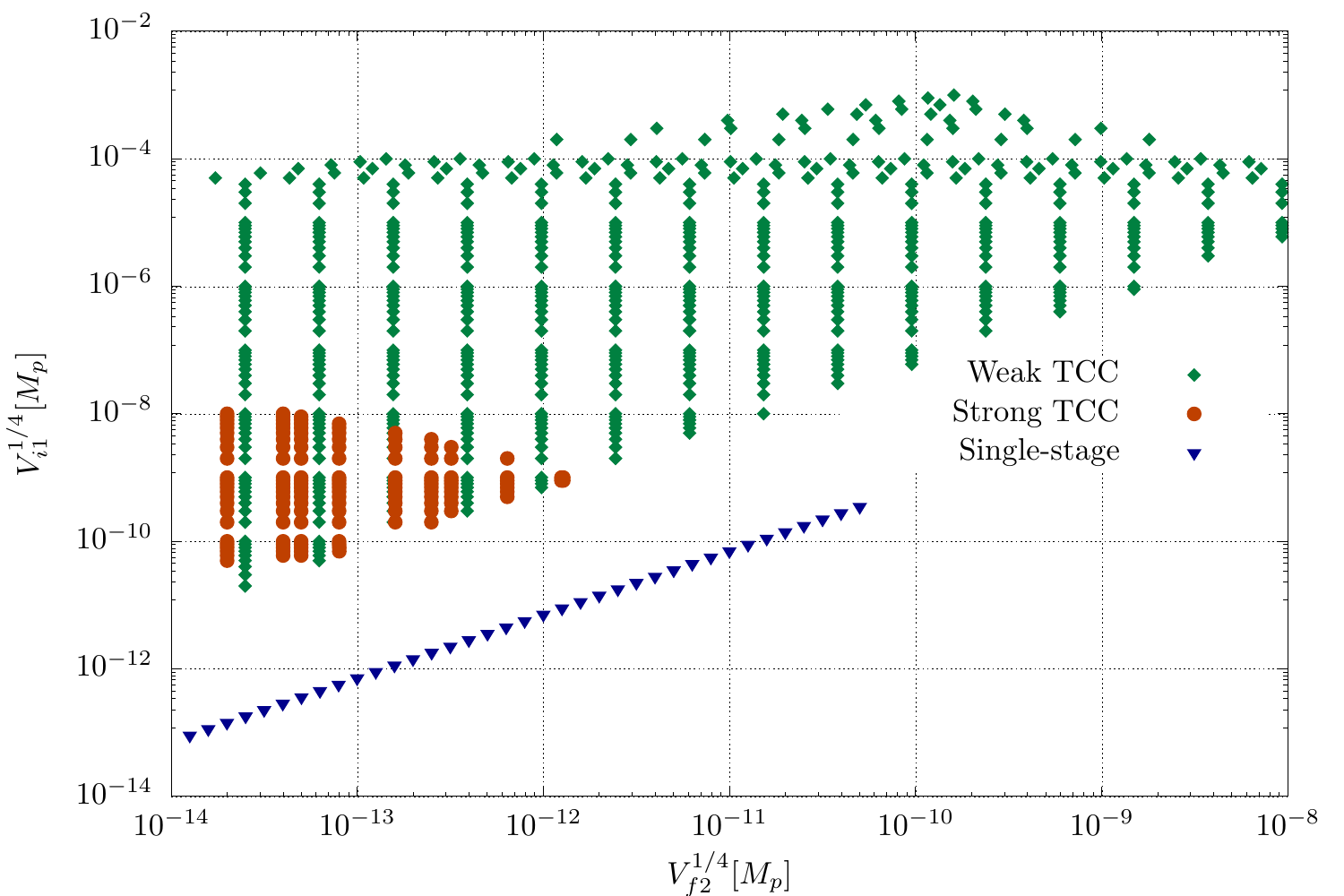}
    {\centering \caption{}}
  \end{subfigure}
~
  \begin{subfigure}[b]{0.49\textwidth}

    \includegraphics[width=1.05\textwidth]{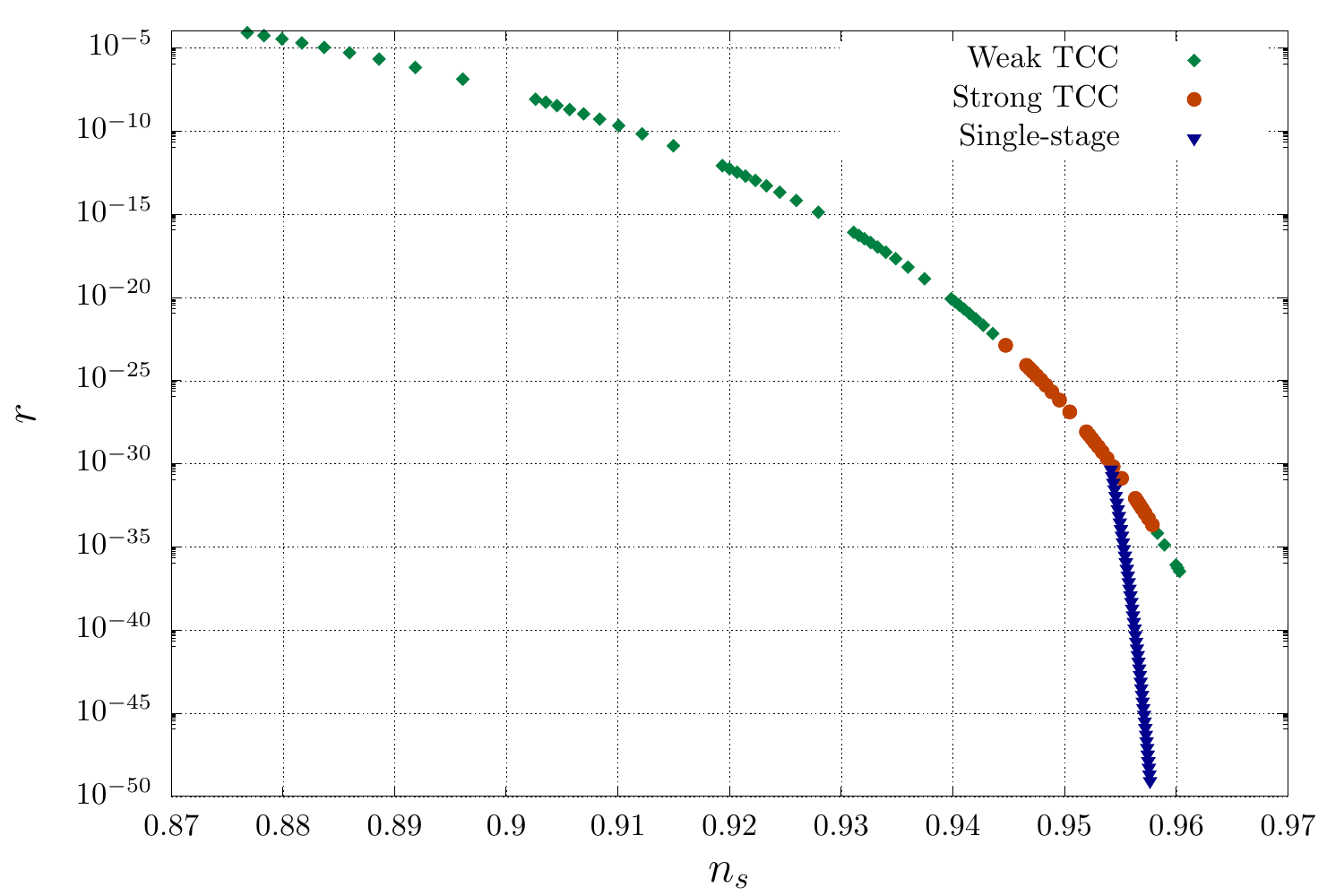}
    \caption{}
  \end{subfigure}
  \caption{Left: Potentials at the beginning, $V_{i1}$, and the end, $V_{f2}$, of inflation that are consistent with the TCC constraints while solving the horizon problem for the toy model presented in Section \ref{toy}. Right: Predictions for the spectral index and tensor-to-scalar ratio for the toy model and for single-stage inflation corresponding to the potentials presented on the left. }
  \label{fig::vp4}
\end{figure*}

Based on the conditions outlined in the previous section, we have sampled different values for the energy scale for the first and second inflationary periods. Since one key goal is to get a higher value of $V_{i1}^{1/4}$, consistent with a non-negligible tensor-to-scalar ratio, this potential is taken as an input. Furthermore, the first expansion is allowed to last as long as possible, i.e., we approach the limit imposed by the TCC by choosing $V_{f1}^{1/4}$ given by Eq. \eqref{vif1}. Then, a good starting point to look for $V_{i2}$ is by exploring the limit set by Eq. \eqref{Nrm}, so that one can simply iterate until the inequality in Eq. \eqref{tcst} holds. An analogous process can be followed for the weak version. Notice that $V_{i2} \sim V_{f2}$ since we are working exclusively in the plateau section of the potential. Thus, once we sample a value of $V_{f2}$ consistent with the imposed constraints, the range of permitted values of $N_2$ can be determined. Finally, using Eqs. \eqref{eqn2} and \eqref{xf2} one can compute $x_{i}$ and $V_{i2}$ for a given value of $N_2$.

Fig. \ref{fig::vp4} (a) shows a set of values for the potential at the end of inflation (corresponding to $V_{f2}$ for the two-stage case) and at the beginning (corresponding to $V_{i1}$ for the two-stage case) which solve the horizon problem while avoiding the crossing of TP modes. Fig. \ref{fig::vp4} (b) presents the corresponding predictions.  Triangular bullets (blue) are used for single-stage warm inflation, while circular (orange) and diamond (green) bullets are used for  two-stage warm inflation considering the strong and the weak TCC, respectively. Below, we will outline how the values presented in the plot can be sampled.

For example, for the strong TCC, take $V_{i1}^{1/4} \sim 10^{-8}\ M_p$, which is two orders of magnitude higher than the allowed values for single-stage inflation. Under this model, the universe can undergo an expansion of $N_1 \sim 41$ e-folds without violating the TCC, which is accomplished with $V_{f1}^{1/4} \sim 1.5 \times 10^{-9}\ M_p$. Thus, the maximum expansion during the intermediate RD phase should be of about $N_R \sim 37$ e-folds, but due to the TCC for the overall expansion history, $V_{i2}$ and $V_{f2}$ are constrained to low values. The highest one can choose, while still having $V_{1} \gg V_2$ is of about $V_{i2}^{1/4} \sim 4 \times 10^{-14}\ M_p$, which renders an expansion during the RD phase of $N_R \sim 10.39$ e-folds. Under these conditions the predicted observables are $n_s \simeq 0.947$ and $r \simeq 8 \times 10^{-25}$, as shown on Fig. \ref{fig::vp4}.

A similar process can be followed for the weak TCC. For instance, take $V_{i1}^{1/4} \sim 10^{-4}\ M_p$. The maximum expansion allowed by the first TCC is $N_1 \sim 22.11$ e-folds, for $V_{f2}^{1/4} \gtrsim 2 \times 10^{-8}\ M_p$. Thus, the maximum expansion during the intermediate RD phase is $N_R \sim 18.62$ e-folds. The more the actual amount of expansion approaches to this value, the wider the range accessible to $N_2$. Take $V_2 \sim 5.4 \times 10^{-10}\ M_p$, which yields an expansion during the RD era of $N_R \sim 10.38$ e-folds. Then, the overall expansion can be roughly between $71$ and $76$ e-folds, or equivalently, $N_2$ can be between $38.51$ and $43.51$ e-folds. The specific value of $N_2$ will determine uniquely $V_{i2}^{1/4}$ and $V_{f2}^{1/4}$. However, notice that they will depart very little from $5.4 \times 10^{-10}\ M_p$. As it can be seen, the weak TCC opens a wider range of accessible values at the energy scales of both inflationary periods, which can be clearly appreciated from Fig. \ref{fig::vp4}.  In this case, the predicted observables are $n_s \simeq 0.90$ and $r \simeq 8.3 \times 10^{-9}$.

Clearly, the predictions shown in Fig. \ref{fig::vp4} (b) and discussed above are model dependent, so one could expect to be able get higher values of $r$ without departing too much from the spectral index measured by the Planck mission ($n_s = 0.9626 \pm 0.0057$) \cite{Akrami:2018odb}. Case in point, for a quadratic potential $V(\phi) \propto \phi^2$ with a constant dissipative coefficient one can get similar constraints as those found for our toy model, although the range of potentials consistent with strong dissipation is more reduced, especially at higher values. In any case, for $V_{i1}^{1/4} \sim 10^{-5}\ M_p$ ($r \sim 10^{-12}$) and $N_1 \simeq 26.9$ e-folds, the quadratic potential model with constant dissipation predicts $n_s\simeq0.977$, which is closer to the experimental value in comparison to $n_s \simeq 0.92$ predicted by the quartic potential with quadratic dissipation.

\section{Discussion}

In this article we have studied the implications of the recently proposed Trans-Planckian censorship conjecture within the warm inflation scenario. We have shown that for a single stage inflation the bounds on the energy scale of inflation are roughly the same as for cold inflation, although there can be some small differences because warm inflation can be realized with steep potentials. However, the bounds on the energy scale, tensor-to-scalar ratio and maximum amount of expansion remain severe. Although theoretically inconvenient, the latter is consistent with the low quadrupole alignment first measured by COBE \cite{Kogut:1996us} and subsequently confirmed by WMAP \cite{Bennett:2012zja} and Planck \cite{Rassat:2014yna}, as shown in \cite{Berera:1997wz}.


Then, following on previous work that suggested the plausibility of periods of warm inflation sourced by cosmological phase transitions, we examined the constraints of the TCC on a two-stage warm inflation scenario with a radiation-dominated era in between. As expected, this kind of models relaxes the amount of inflation required during each stage in order to solve the horizon problem, which enhances the range of allowed energies, while still being consistent with the TCC. Our toy model showed that one can access energies up to two orders of magnitude higher than those allowed for a single-stage scenario. Naturally, these bounds could be improved by further increasing the number of inflationary stages.

In another direction, we also analyzed the possibility of having two (or more) independent meta-stable dS states, leading to a weaker version of the TCC. In a broader sense, such states could be seen as emerging from different local regions sampling an infinite distance in moduli space (see Ref. \cite{Draper:2019utz}). Furthermore, considering that the TCC could be a consequence of the swampland distance conjecture \cite{Brahma:2019vpl}, the weak TCC could follow.
With these considerations, our toy model showed that the tensor-to-scalar ratio could be as high as $10^{-5}$, although it corresponds to red spectral indices. However, this particular prediction is highly model dependent, so it would be interesting to further explore this idea by considering more realistic dissipative coefficients, as those obtained from quantum field theory calculations.

In conclusion, if proven to be correct, the TCC puts stringent constraints on inflationary models coupled to gravity. Alternatively, one could argue that the TCC in and of itself provides a recipe to avoid the technical and conceptual inconvenience of having TP modes crossing the horizon, and thus to obtain predictions without relying on unknown high-energy physics.
In any case, scenarios like warm inflation may generically satisfy and/or avoid those constraints while also predicting tensor perturbations that could be measured in future experiments. In addition to this, it can also accommodate other setups, like having a thermal bath of gravitons, which could also enhance the amplitude of tensor perturbations.


\begin{acknowledgements}
AB is supported by STFC. JRC is supported by the Secretary of Higher Education, Science, Technology and Innovation of Ecuador (SENESCYT).
\end{acknowledgements}

\bibliography{apssamp}

\end{document}